\documentclass[conference]{IEEEtran}

\usepackage{lineno,hyperref}
\usepackage{csquotes}

\usepackage[]{todonotes}

\usepackage{rotating}

\usepackage{multirow}

\usepackage{subcaption}

\usepackage{paralist}

\usepackage{fontawesome}

\usepackage[capitalise,nameinlink]{cleveref}
\crefname{section}{Sect.}{Sect.}
\Crefname{section}{Section}{Sections}
\crefname{figure}{Fig.}{Figs.}
\Crefname{figure}{Figure}{Figures}
\crefname{table}{Tab.}{Tabs.}
\Crefname{table}{Table}{Tables}
\crefname{lstlisting}{List.}{List.}
\Crefname{lstlisting}{Listing}{Listings}

\usepackage[inline]{enumitem}

\newcommand{\eg}{e.\,g.,\ }
\newcommand{\ie}{i.\,e.,\ }

\begin{document}

\title{GraphScale: Scalable Bandwidth-Efficient Graph Processing on FPGAs}

\author{\IEEEauthorblockN{Jonas Dann}
\IEEEauthorblockA{SAP SE and
Heidelberg University\\
Walldorf, Germany\\
jonas.dann@sap.com}
\and
\IEEEauthorblockN{Daniel Ritter}
\IEEEauthorblockA{SAP SE\\
Walldorf, Germany\\
daniel.ritter@sap.com}
\and
\IEEEauthorblockN{Holger Fröning}
\IEEEauthorblockA{Heidelberg University\\
Heidelberg, Germany\\
holger.froening@ziti.uni-heidelberg.de}}

\maketitle

\begin{abstract}
Recent advances in graph processing on FPGAs promise to alleviate performance bottlenecks with irregular memory access patterns.
Such bottlenecks challenge performance for a growing number of important application areas like machine learning and data analytics.
While FPGAs denote a promising solution through flexible memory hierarchies and massive parallelism, we argue that current graph processing accelerators either use the off-chip memory bandwidth inefficiently or do not scale well across memory channels.

In this work, we propose GraphScale, a scalable graph processing framework for FPGAs.
For the first time, GraphScale combines multi-channel memory with asynchronous graph processing (\ie for fast convergence on results) and a compressed graph representation (\ie for efficient usage of memory bandwidth and reduced memory footprint).
GraphScale solves common graph problems like breadth-first search, PageRank, and weakly-connected components through modular user-defined functions, a novel two-dimensional partitioning scheme, and a high-performance two-level crossbar design.
\end{abstract}

\section{Introduction}
\label{sec:intro}
\begin{figure}[bt]
    \centering
    \includegraphics[width=\linewidth]{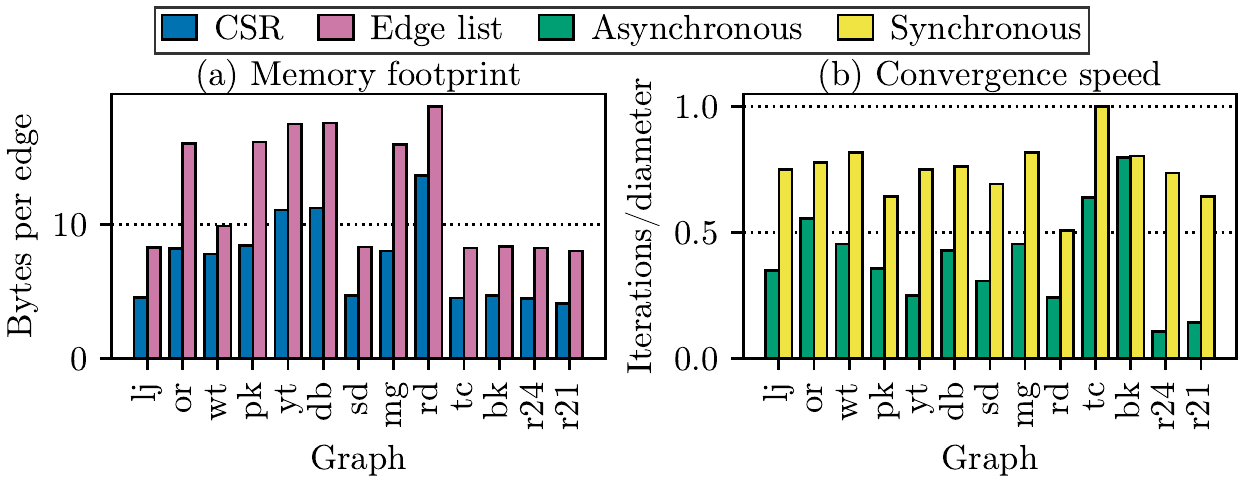}
    \vspace{-1.75em}
    \caption{Memory footprint and convergence speed for asynchronous graph processing on a compressed data structure}
    \label{fig:appetizer}
    \vspace{-1em}
\end{figure}

Irregular memory access and little computational intensity inherent to graph processing cause major performance challenges on traditional hardware (\eg CPUs) \cite{journals/corr/abs-1910-09017, journals/corr/abs-2007-07595, LumsdaineGHB07, conf/isca/AhnHYMC15}.
Field programmable gate arrays (FPGAs) promise to accelerate common graph problems like breadth-first search (BFS), PageRank (PR), and weakly-connected components (WCC) with their flexible memory hierarchy (\eg by low-latency on-chip memory) and massive parallelism \cite{journals/corr/abs-1910-09017}.
Still, memory bandwidth is the bottleneck of graph processing even for highly optimized FPGA implementations.
Thus, graph processing accelerators like AccuGraph \cite{conf/IEEEpact/Yao0L0H18} and FabGraph \cite{conf/fpga/ShaoLHL019} utilize \emph{graph compression} and \emph{asynchronous graph processing} to reduce the load on the memory sub-system.
\cref{fig:appetizer} shows the potential of these two graph processing accelerator properties.
For graphs with large average degree, a well-known compressed graph data structure like compressed sparse row (CSR) almost halves the number of Bytes per edge to be processed.
Asynchronous graph processing, in turn, may lead to a significant decrease in iterations over the graph.
However, these approaches have not yet been scaled to multiple memory channels limiting their performance on modern hardware \cite{conf/sigmod/Dann0F21}.

In this paper, we introduce GraphScale, the first scalable graph processing framework for FPGAs based on asynchronous graph processing on a compressed graph.
GraphScale competes with other scalable graph processing frameworks that work on labeled graphs, like HitGraph \cite{journals/tpds/ZhouKPSW19} and ThunderGP \cite{journal/trets/ChenCTCHWC22} that, however, do not leverage the potential shown in \cref{fig:appetizer}.
While, for asynchronous graph processing, the challenge lies in handling the high-bandwidth data flow of vertex label reads and writes to on-chip scratch pads at scale, the CSR-compressed graph adds design complexity and higher resource utilization for materializing compressed edges on-chip that is challenging for scaling to multiple memory channels.

To leverage combined asynchronous graph processing and graph compression, we make the following contributions:
\begin{itemize}
    \item We design an asynchronous graph framework, efficiently solving common graph problems on compressed data.
    \item We design a scalable, resource-efficient two-level vertex label crossbar.
    \item We propose a novel two-dimensional partitioning scheme for graph processing on multiple memory channels.
    \item We evaluate our approach compared to state-of-the-art, scalable graph processors: HitGraph and ThunderGP.
\end{itemize}

The resulting GraphScale system shows promising scalability with a maximum speedup of $4.77\times$ on dense graphs and an average speedup of $2.3\times$ over HitGraph and ThunderGP.
Overall, we conjecture that asynchronous processing on a compressed graph with multi-channel memory scaling improves the graph processing performance, however, leads to interesting trade-offs (\eg for large graphs where partitioning overhead dominates performance).

\section{Background and Related Work}
\label{sec:background}
In this section, we introduce data structures, implementation schemes, and graph problems important to graph processing on FPGAs.
Additionally, we discuss related work and our contributions in the context of current graph accelerators.

\subsection{Graph Processing}
\label{sec:graph}
\begin{figure}[bt]
    \centering
    \includegraphics[width=.75\linewidth]{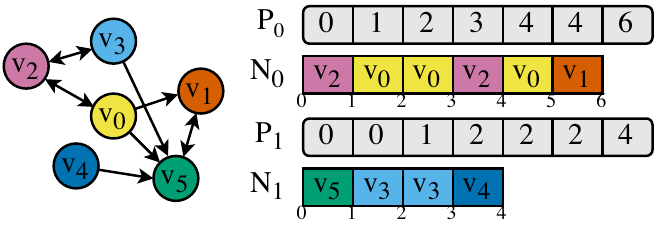}
    \caption{Directed graph and corresponding inverse horizontally-partitioned CSR (adapted from \cite{conf/btw/Dann0F21})}
    \label{fig:graph}
    \vspace{-0.5em}
\end{figure}
Graphs are abstract data structures ($G = (V,E)$) comprising a vertex set $V$ and an edge set $E \subseteq V \times V$.
Current graph processing accelerators represent the graphs they are working on in memory either as an array of edges also called edge list or a set of adjacency lists, each containing a vertex's neighbors in the graph.
One possible implementation of an adjacency lists structure is compressed sparse row (CSR), which we use in GraphScale.
To allow for more meaningful graph processing, a label is attached to each vertex, unlike \cite{conf/fpga/ZhangL18} and ScalaBFS \cite{conf/fpga/LiuSLWCLL021}.
Additionally, graph processing accelerators utilize two dimensions of partitioning: vertical and horizontal.
Vertical partitioning divides the vertex set into intervals such that each partition contains the incoming edges of one interval.
In contrast, horizontal partitioning again divides up the vertex set into intervals but each partition contains the outgoing edges of one interval.
In GraphScale, we horizontal partition the set of inverse edges (the result for an example graph is shown in \cref{fig:graph}) and extend it to work on multiple memory channels.
The values of the pointers array (P) at position $i$ and $i{+}1$ delimit the neighbors (N) of $v_i$.
For example, for vertex $v_5$ in partition 1 these are the values of the neighbors array between $2$ and $4$ (\ie $v_3$ and $v_4$).
As a third partitioning approach, interval-shard partitioning \cite{conf/usenix/ZhuHC15} employs both vertical and horizontal partitioning at once.

Depending on the underlying graph data structure, graphs are processed based on two fundamentally different approaches: edge- and vertex-centric graph processing.
Edge-centric systems iterate over the edges as primitives of the graph on an underlying edge list.
Vertex-centric systems iterate over the vertices and their neighbors as primitives of the graph on an underlying adjacency lists data structure (\eg CSR).
The vertex-centric approach can be further divided into push- and pull-based data flow.
A push-based data flow denotes that updates are pushed along the forward direction of edges to update neighboring vertices, while in a pull-based data flow updates are pulled along the inverse direction of edges from neighboring vertices to update the current vertex.
Lastly, there are currently two dominant update propagation schemes.
Asynchronous graph processing directly applies updates to the working vertex label set when they are produced and synchronous graph processing collects all updates in memory and applies them only after the iteration is finished.

In the context of this work, we consider the three graph problems BFS, PR, and WCC.
BFS denotes a sequence of visiting the vertices of a graph.
Vertices are labeled with their distance (in length of the shortest path in number of edges) to a root vertex.
WCC specifies as output for each vertex its affiliation to a weakly-connected component.
Two vertices are in the same weakly-connected component, if there is an undirected path between them.
PR is a measure to describe the importance of vertices in a graph.
It is calculated as $p(i, t{+}1) = \frac{1 - d}{|V|} + d \cdot \sum_{j \in N_G(i)} \frac{p(j, t)}{d_G(j)}$ for each $i \in V$ with damping factor $d$, neighbors $N_G$, degree $d_G$, and iteration $t$.

\subsection{Related Work}
\label{sec:relatedwork}
\begin{table*}[t]
\caption{Related FPGA-based graph processing accelerators with classification and feature set}
\label{tab:systems}
\footnotesize
\centering
\begin{tabular}{l|l l l l|c c c|c c}
    Identifier & Iteration scheme & Flow & Partitioning & Data structure & Compressed & Async. & Scales & Labels & Framework \\
    \hline
    \hline
    AccuGraph \cite{conf/IEEEpact/Yao0L0H18} & Vertex-centric & Pull & Horizontal & inverse-CSR & \faThumbsUp & \faThumbsUp & \faThumbsODown & \faThumbsUp & \faThumbsUp \\
    FabGraph \cite{conf/fpga/ShaoLHL019} & Edge-centric & n/a & Interval-shard & Compr. edge list & \faThumbsUp & \faThumbsUp & \faThumbsODown & \faThumbsUp & \faThumbsUp \\
    HitGraph \cite{journals/tpds/ZhouKPSW19} & Edge-centric & n/a & Horizontal & Sorted edge list & \faThumbsODown & \faThumbsODown & \faThumbsUp & \faThumbsUp & \faThumbsUp \\
    ThunderGP \cite{journal/trets/ChenCTCHWC22} & Edge-centric & n/a & Vertical & Sorted edge list & \faThumbsODown & \faThumbsODown & \faThumbsUp & \faThumbsUp & \faThumbsUp \\
    \hline
    Zhang et al. \cite{conf/fpga/ZhangL18} & Vertex-centric & Hybrid & None & CSR & \faThumbsUp & \faThumbsODown & \faThumbsUp & \faThumbsODown & \faThumbsODown \\
    ScalaBFS \cite{conf/fpga/LiuSLWCLL021} & Vertex-centric & Hybrid & None & CSR & \faThumbsUp & \faThumbsODown & \faThumbsUp & \faThumbsODown & \faThumbsODown \\
    \hline
    \textbf{GraphScale} & Vertex-centric & Pull & Custom & inverse-CSR & \faThumbsUp & \faThumbsUp & \faThumbsUp & \faThumbsUp & \faThumbsUp \\
\end{tabular}

\medskip
Compressed: Compressed data structure; Async.: Asynchronous graph processing; n/a: Not applicable; \faThumbsUp: yes, \faThumbsODown: no
\end{table*}

\Cref{tab:systems} shows a representative sub-set of current FPGA-based graph processing accelerators, found in a recent survey \cite{journals/corr/abs-2007-07595}, and how they relate to GraphScale.
AccuGraph \cite{conf/IEEEpact/Yao0L0H18}, FabGraph \cite{conf/fpga/ShaoLHL019}, HitGraph \cite{journals/tpds/ZhouKPSW19}, and ThunderGP \cite{journal/trets/ChenCTCHWC22,conf/fpga/ChenTCHWC21} all work on labeled graphs and are frameworks that map to multiple graph problems.
The latter three employ edge-centric graph processing with different partitioning schemes.
HitGraph and ThunderGP sort their edge list beforehand to enable update coalescing before writing updates back into memory and applying them in a second phase in each iteration (\ie both synchronous).
FabGraph, in contrast, employs interval-shard partitioning, which enables compression of vertex identifiers for each partition and asynchronous update propagation.
AccuGraph follows a vertex-centric approach with a pull-based data flow on a inverse, horizontally-partitioned CSR data structure.
While FabGraph and AccuGraph enable the potential of a compressed data structure and asynchronous graph processing, they do not scale to multiple memory channels and HitGraph and ThunderGP scale to multiple memory channels but do not exploit that potential.
GraphScale combines both sides on a conceptual basis similar to AccuGraph.

Besides these frameworks, there are specialized graph processing accelerators---surpassing the performance of general-purpose frameworks due to the much simpler memory access pattern and often massive memory bandwidth of their benchmark systems---that do not map to other graph problems and do not work on labeled graphs.
For example, Zhou et al. \cite{conf/reconfig/ZhouCP15} propose a PR accelerator, while dedicated BFS accelerators were proposed by \cite{conf/asap/BetkaouiWTL12, conf/ipps/AttiaJTJZ14, journals/iracst/LeiRG15} on the Convey HC-2 system, which is not commercially available anymore.
Zhang et al. \cite{conf/fpga/ZhangL18} uses high-bandwidth hybrid memory cube (HMC), allowing for extreme performance scaling, however, through BFS-specific simplifications and assumptions.
Similarly, ScalaBFS \cite{conf/fpga/LiuSLWCLL021} reports superior performance on modern high-bandwidth memory (HBM), but is also limited to BFS.
Notably, due to the problem-specific simplifications, those accelerators cannot easily be extended to support other graph problems (\eg simply using BFS for WCC is inefficient).

\section{GraphScale System Design / Architecture}
\label{sec:concept}
\begin{figure}[bt]
    \centering
    \includegraphics[width=.8\linewidth]{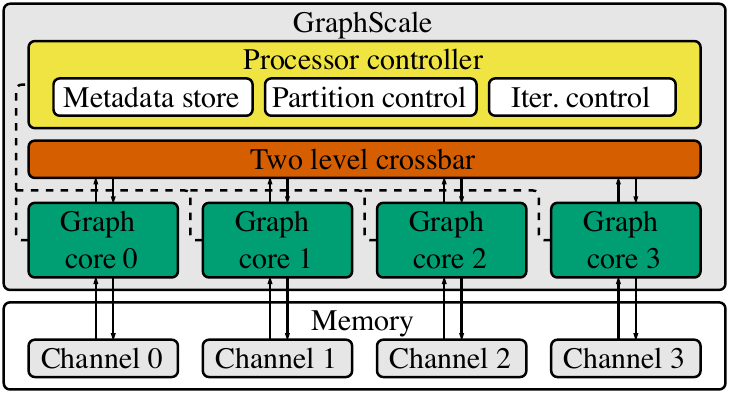}
    \caption{GraphScale overview}
    \label{fig:system}
\end{figure}
In this section, we describe our graph processing framework GraphScale (cf. \cref{fig:system}) at an abstract processor-scale level and subsequently explain its non-trivial design concepts.
In principle, a GraphScale graph processor consists of $p$ graph cores (explained in \cref{sec:core}) matched to the number of $p$ memory channels (four in this example).
Each graph core is only connected to its own memory channel and can thus only directly read and write data on this channel.
This requires partitioning of the graph into at least $p$ partitions.
The details of partitioning and how the graph is distributed over the memory channels is discussed in \cref{sec:partitioning}.
However, since graph partitioning does not eliminate data dependencies between partitions, the graph cores are connected via a high-performance crossbar for exchange of vertex labels enabling the scaling of the approach.
The crossbar will be explained in \cref{sec:crossbar}.
The whole execution is governed by a processor controller.
Before execution starts, the host code passes parameters for each partition and optimization flags to the processor controller which stores them in a metadata store.
When execution is triggered by the host code, the processor goes through a state machine, orchestrating the control signals for the execution of iterations over the graph.

\subsection{Graph Core}
\label{sec:core}
\begin{figure*}[bt]
    \centering
    \includegraphics[width=.75\linewidth]{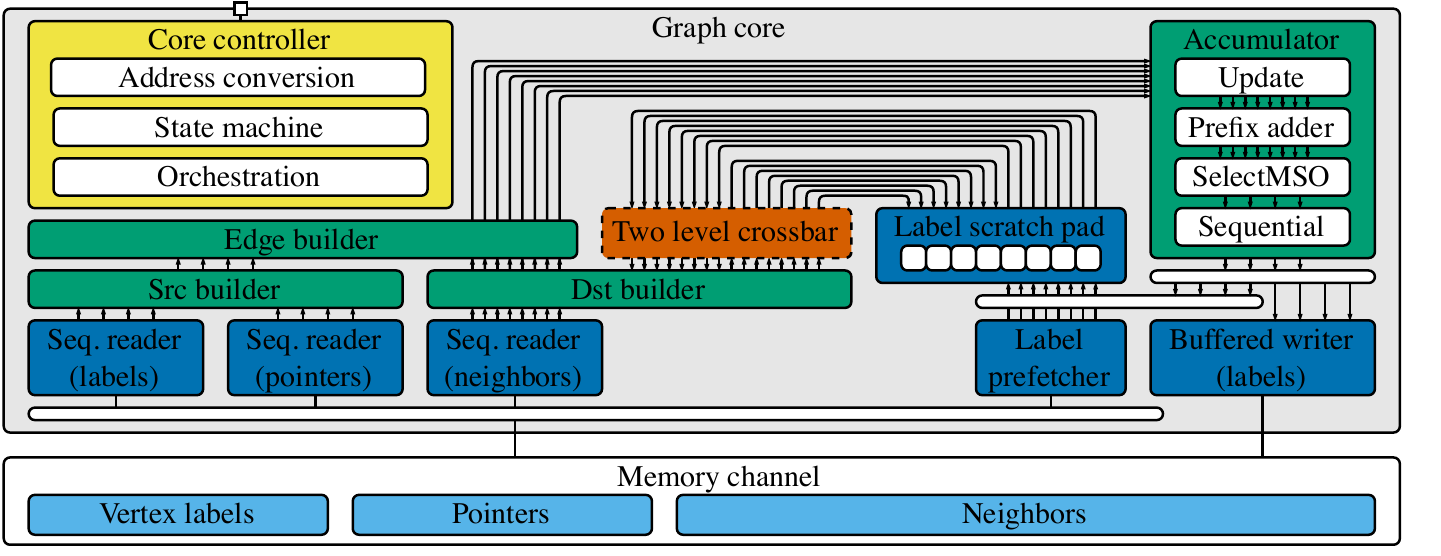}
    \caption{Graph core overview}
    \label{fig:core}
\end{figure*}
A graph core (cf. \cref{fig:core}), as the basic building block of GraphScale, processes graphs based on the vertex-centric iteration scheme and pull-based data flow.
It works on a partitioned inverse-CSR data structure of the graph (cf. \cref{sec:graph}) consisting of one vertex labels array and for each partition one pointers and one neighbors array.
Furthermore, processing of the graph is structured into two phases per iteration: prefetching and processing.
In the prefetching phase, the vertex label prefetcher reads a partition specific interval of the vertex label array into the label scratch pad, an on-chip memory (BRAM) split up into $e$ (set to $8$ in this example) banks.
The label scratch pad is used to serve all non-sequential read requests that occur during an iteration instead of off-chip DRAM since BRAM has much higher bandwidth and predictable one cycle request latency independent of the access pattern.

Starting the data flow of the processing phase, the source builder reads vertex labels and pointers sequentially.
Vertex labels and pointers are zipped to form $v$ source vertices in parallel with a vertex index (generated on-the-fly), vertex label, inclusive left bound, and exclusive right bound of neighbors in the neighbors array each.
The destination builder reads the neighbors array of the current partition sequentially and puts $e$ neighbor vertex identifiers in parallel through the two level crossbar passing the vertex identifiers to the correct label scratch pad bank of the correct graph core and returning the resulting vertex labels in the original order (discussed in more detail in \cref{sec:crossbar}).
The vertex label annotated with the neighbor index is then passed to the edge builder which combines source and destination vertices based on the left bound $l$ and right bound $r$ of the source vertex and the neighbor index $j$ of the destination vertex as $l <= j; j < r$.
Thus, we get up to $e$ edges with a maximum of $v$ source vertices as output from the edge builder per clock cycle.

The accumulator takes the $e$ edges annotated with their source and destination vertex labels as input and updates vertices in four steps.
First, updates are produced in the Update stage depending on the graph problems' update function for each edge in parallel. 
For BFS, this means taking the minimum of the source vertex label and destination vertex label plus $1$.
If the latter is smaller, the output is flagged as an actual update of the source vertex label.
This is crucial for algorithms that terminate when no more updates are produced in an iteration (\eg BFS and WCC).
The pairs of source vertex identifier and updated vertex labels are then passed to the Prefix Adder which reduces the updates to the most significant element with the same source vertex identifier for each source vertex.
The most significant entry is then selected by the $v$ selectors in the SelectMSO stage of the accumulator and passed on to the last stage.
Each selector already only selects pairs with $i \% v = I$ for source vertex index $i$ and selector index $I$.
The Sequential stage consists of $v$ sequential operators that reduce updates from subsequent cycles to the same vertex into one that is output when a new source vertex identifier is encountered or a higher source vertex identifier is encountered.
Thus, in total, the accumulator produces updates only when the new label is different based on the annotated edges and reduces them into a maximum of one update per source vertex.

\begin{figure}[bt]
    \centering
    \includegraphics[width=\linewidth]{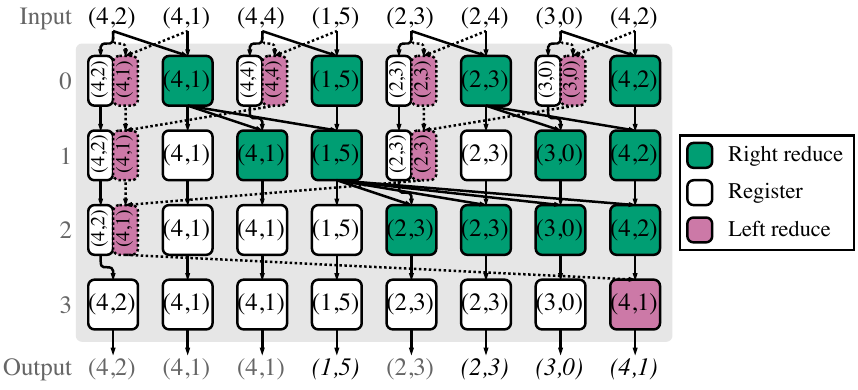}
    \caption{Prefix adder with wrap-around suffix sub-accumulator (with minimum reduce operator)}
    \label{fig:accumulator}
\end{figure}
\cref{fig:accumulator} shows the parallel prefix-adder vertex-update accumulator.
We add a suffix sub-accumulator (dotted outlines) necessary to attain correct results in some edge cases and a merged signal for non-idempotent reduce operators like summation.
The accumulator takes $e$ pairs of source vertex identifier and updated vertex label (split with a comma) and returns one updated vertex label as the right-most identifier-label pair per incoming source vertex identifier (italicized).
The prefix-adder accumulator consists of $\log_2(e) + 1$ pipelined levels of registers (white) and reduce processing elements (PE).
The registers take one identifier-label pair as an input and passes this input on in the next clock cycle.
The reduce PEs (green and pink) take two identifier-label pairs as an input and combine them depending on the graph problem the graph core maps if the source vertex identifiers are equal.
The result is again put out in the next clock cycle.
Right reduce PEs (green) pass on the right identifier-label pair unmodified if the identifiers are unequal and left reduce PEs (left) pass on the left pair.
In this particular example, the parallel accumulator could either be used \eg for BFS or WCC because it uses minimum reduce PEs which put out the minimum of the vertex labels if they should be combined.
The connection pattern of the first $\log_2 e$ levels of the accumulator represent a Ladner-Fischer prefix-adder.

Additional to the prefix adder, we also introduce a suffix sub-adder which reduces all identifier-label pairs with source vertex identifier equal to the first one to the first element.
In an additional pipeline step, this suffix accumulation result is reduced with the last prefix accumulation result if there have been multiple different source vertex identifiers in the input.
We do this because the sequential source vertex identifiers can overlap from the last one to the first one as a result of how the edge builder works.
In this edge case updated vertex labels might be missed because only the right-most vertex label of a source vertex identifier is further processed.
Finally, we only reduce two identifier-label pairs if all pairs in between have the same source vertex identifier which we keep track of with a merged signal mentioned above.

The resulting updates are fed back to a buffered writer and into the label scratch pad so they can immediately be used in the same iteration.
The buffered writer collects all updates to the same cache line and writes them back to memory when an update to a new cache line is encountered.

All different parts of this design are orchestrated in their execution by a core controller.
The core controller gets graph-wide parameters of the execution like number of vertices, number of edges and address of the buffer in memory and dynamic parameters like iteration number from the processor controller.
Based on this, it starts the prefetch phase and then the processing phase and therefore calculates the addresses to the data structure arrays.
Finally it also flushes the pipeline so all updates are written back to memory before asserting the ready signal such that the next iteration can be started.

\subsection{Scaling Graph Cores}
\label{sec:crossbar}
To scale this very effective single-channel design with as limited overhead as possible, we propose the graph core memory channel assignment shown already in \cref{fig:system}.
Each graph core works on exactly one memory channel.
However, requests to vertex labels require communication between graph cores. 
Therefore, we propose a scalable resource-efficient two level crossbar.
In this section, we will describe how we achieve the necessary high throughput of this crossbar to saturate the accumulators of multiple graph cores with annotated edges.

\begin{figure}[bt]
    \centering
    \includegraphics[width=.9\linewidth]{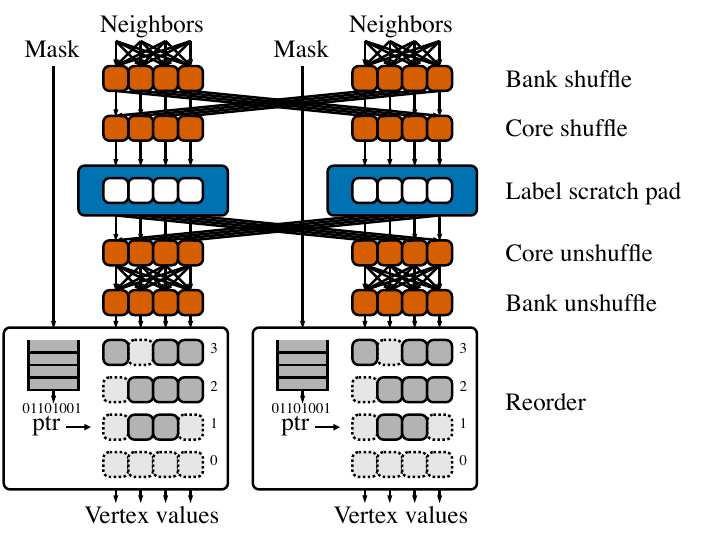}
    \vspace{-.1cm}    
    \caption{Stages of scalable two level crossbar}
    \label{fig:shuffle-reorder}
    \vspace{-.5cm}
\end{figure}
We show the full shuffle reorder flow of the crossbar for two cores and $e = 4$ in \cref{fig:shuffle-reorder}.
The first level (bank shuffle) of this flow gets $e$ neighbors from the destination builder each cycle for each core.
The neighbor indices serve as addresses to vertex labels in the labels array.
Before the processing of a partition starts, the partition's vertex label are prefetched to the label scratch pad to serve these requests.
Since memory returns $e$ neighbors per graph core per cycle at maximum throughput, $e * p$ requests have to be served by the label scratch pad per cycle.
Thus, the label scratch pad of each graph core is split up into $e$ banks that can serve requests in parallel and the vertex labels are striped over these banks.
This means that the last $\log_2 e$ bits of the neighbor index are used to address the bank of the label scratch pad that this vertex label can be requested from.
Thus, the bank shuffle level puts each neighbor index into the right bank lane based on its last $\log_2 e$ bits.
This can introduce stalls because multiple neighbors from one line can go to the same bank (\ie a multiplexer has to output entries from the same input cycle in multiple output cycles).
However, since each neighbor only goes to one bank we decouple the $e$ bank shufflers and let them consume multiple full lines before stalling such that labels from later lines can overtake in other banks.
For most graphs, this reduces stalls because the load is approximately balanced between banks.

In a second level, we introduce a core crossbar that shuffles the neighbor indices annotated with their line and lane they originally came from to the core that contains the vertex label.
Core addressing is done by the first $\log_2 p$ bits of the neighbor index.
However, since the neighbor indices are already in the correct lane, this only requires $p * e$ core shufflers with $p$ inputs.
The results are additionally annotated with the core they originally came from and fed into the label scratch pad.
The core shufflers also work independently from each other, allowing neighbor indices to overtake each other.

The label scratch pad returns the vertex labels with a one cycle latency but keeps the annotations.
A second layer of core shufflers routes the vertex labels back to their original graph core.
Thereafter, the vertex labels are unshuffled to the lane they originally came from and fed into a final reorder stage which is necessary to restore the original sequence of the data which is possibly changed because requests and responses overtake each other in the previous steps.

The reorder stage has a fixed number of lines it can keep open at a time ($4$ in this example) which we will call reorder slots.
It is passed the valid signals of the incoming neighbors when they first enter the crossbar and puts them in a FIFO.
The unshuffled labels are then still annotated with the line they originally came from modulo the number of reorder slots which is used as the address to put them in a BRAM.
There is one BRAM for each lane of reorder slots because in each cycle we possibly write one label and read one label per lane.
The reorder stage also maintains a pointer pointing to the currently to be output line and compares the valid signals of this line to the FIFO output.
If the FIFO valid output and the valid signals form the line are equal, the labels are put out, the pointer is incremented, the FIFO is popped, and the valid signals of the line are cleared.
When the pointer is incremented above the last line it overflows to 0.

Finally, the FIFO of the reorder stage is also used to exert backpressure.
If the FIFO has as many elements as there are reorder slots, the ready signal is deasserted and all stages stop.
To handle the one cycle latency of the label scratch pad there is also an additional overflow register where the label scratch pad result can overflow to.

\subsection{Graph Partitioning and Optimization}
\label{sec:partitioning}
\begin{figure}[bt]
    \centering
    \includegraphics[width=\linewidth]{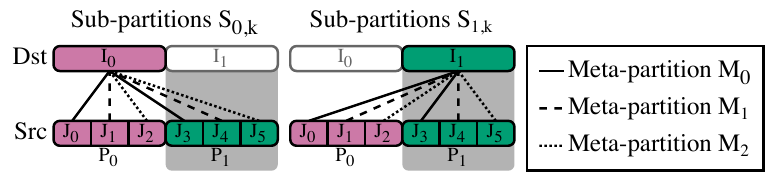}
    \caption{Two-dimensional partitioning (two cores and three sub-intervals)}
    \label{fig:partitioning}
\end{figure}
\cref{fig:partitioning} shows the partitioning of the input graph as the last missing part of our graph processing accelerator GraphScale and why it is able to scale well.
The partitioning in done in two dimensions.
In the first dimensions, the set of vertices is divided into $p$ equal intervals $I_q$ ($I_0$ and $I_1$ in this examples for $p=2$), one stored on each memory channel and processed by its corresponding graph core $P_q$.
The second dimension of partitioning divides each vertex interval into $l$ equal sub-intervals ($J_0$ to $J_5$ in this example for $l=3$) that fit into the label scratch pad of the graph core.
We generate one sub-partition $S_{i,j}$ for each pair of interval $I_i$ and sub-interval $J_j$ containing all edges with destination vertices in $I_i$ and source vertices in $J_j$ and rewrite the neighbor indices in the resulting CSR data structure such that the requests are shuffled to the correct graph core by the two level crossbar (\ie first $\log_2 p$ bits are graph core index) and subtract the offset of the respective sub-interval $J_j$.
Sub-partitions $S_{q,q*l}$ for each $q \in [0,q)$ additionally form a meta-partition $M_q$.
During execution, all sub-intervals $J_{q*l}$ are prefetched by their respective graph core $q$ before processing of all sub-partitions of meta-partition $M_q$ is triggered.
This light-weight graph partitioning, however, may introduce load imbalance because it works on simple intervals of the vertex set.

Each graph core writes all label updates to off-chip memory through the buffered writer while processing a partition.
As a first optimization, immediate updates, we also immediately write back the updates to the vertex label scratch pad if they are part of the current partition \cite{conf/IEEEpact/Yao0L0H18}.
Thus, with this optimization, BRAM and off-chip memory are always in sync.
Nevertheless, at the beginning of processing a partition, the vertex label set is unnecessarily prefetched even if the labels are already present in BRAM.
Thus, we utilize prefetch skipping in this case as a light-weight control flow optimization which skips the prefetch phase if the vertex label set is already present in the label scratch pad \cite{conf/btw/Dann0F21}.
This optimization only works in conjunction with immediate updates.
As a third optimization, we apply stride mapping to improve partition balance which we identified as a large issue during testing \cite{conf/fpga/DaiHCXWY17}.
Because the graph cores work in lock-step on the meta-partitions, imbalanced partitions lead to a lot of idle time.
Stride mapping is a light-weight technique for semi-random shuffling of the vertex identifiers before applying the partitioning which creates a new vertex ordering with a constant stride ($100$ in our case which results in $v_0, v_{100}, v_{200}, ...$).

\section{Evaluation}
\label{sec:evaluation}
In this section, we introduce the system used for evaluation and setup for measuring performance together with the system parameters and the graph data sets used for the experiments.
Thereafter, we comprehensively measure performance in multiple dimensions.
We first look at the effects of the optimizations introduced in \cref{sec:optimizations}, before highlighting the scalability of the GraphScale framework.
Lastly, we will compare GraphScale with the performance of the competitors that scale to multiple memory channels (\ie HitGraph and ThunderGP) to show the advantages and disadvantages of a compressed graph and asynchronous graph processing.

\begin{figure}[bt]
    \centering
    \includegraphics[width=.9\linewidth]{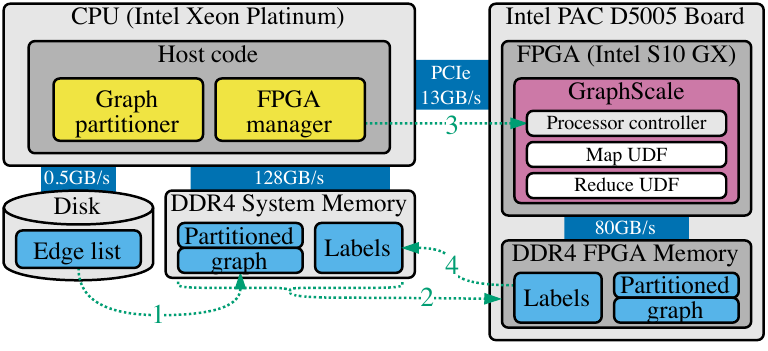}
    \caption{Overall system architecture (incl. host, device, memory)}
    \label{fig:computer-architecture}
\end{figure}
\cref{fig:computer-architecture} shows the system context in which GraphScale is deployed.
In principle, the system features a CPU and an accelerator board which hosts the FPGA, running GraphScale itself, and memory, used as intermediate data storage for the graph during processing.
The CPU manages the execution on the FPGA and is also responsible for loading and partitioning the graph.
To execute a particular workload with a particular graph, the GraphScale framework, first, is synthesized with user defined functions (UDFs) for the map and reduce operators in the graph core accumulator.
Map produces an updates to the source vertex label for each edge, while reduce aggregates updates into one value for each to be updated vertex.
For a switch from BFS to WCC, the reduce UDF stays the same while only one line has to be changed for the map UDF.
PR requires changing the map UDF significantly and replacing the reduce UDF with summation.
Additionally, PR alternatingly works on two separate vertex label arrays.
Secondly, the synthesized design is programmed to the FPGA.

For execution of the programmed algorithm on a particular graph data set, the edge list (or any other representation) of the graph is read from disk to the CPU and partitioned by the graph partitioner in the host code according to the GraphScale framework parameters.
Additionally, the vertex labels of the graph are initialized with graph problem specific values.
The graph partitions and vertex labels are then transferred to the respective channels of the FPGA memory.
Thereafter, the parameters of the graph are passed to GraphScale via a control interface which triggers execution.
After the execution finished, the results can be read back to CPU memory and used for further processing.
If desired, the partitioned graph can be used multiple times in a row by loading new vertex labels and again triggering the execution.

For our experiments, we are working with a server equipped with an Intel FPGA Programmable Accelerator Card (PAC) D5005 attached via PCIe version 3.
The system features two Intel Xeon Gold 6142 CPUs at 2.6GHz and 384GB of DDR4-2666 memory.
The D5005 board is equipped with 4 channels of DDR4-2400 memory with a total capacity of 32GB and a resulting bandwidth of $76.8$GB/s.
The design itself is based on the Intel Open Programmable Execution Engine (OPAE) platform and is synthesized with Quartus 19.4.

\begin{table}[bt]
\caption{Resource utilization and clock frequency by graph problem and number of graph cores}
\label{tab:utilization}
\footnotesize
\centering
\begin{tabular}{l r|r r r r r r}
 Problem & $p$ & LUTs & Regs. & BRAM & DSPs & Clock freq. \\
 \hline
 \hline
 BFS & $1$ & $19$\% & $13$\% & $40$\% & $0$\% & $192$MHz \\
 BFS & $2$ & $30$\% & $23$\% & $41$\% & $0$\% & $186$MHz \\
 BFS & $4$ & $58$\% & $47$\% & $43$\% & $0$\% & $170$MHz \\
 \hline
 PR & $1$ & $26$\% & $14$\% & $66$\% & ${<}1$\% & $174$MHz \\
 PR & $2$ & $43$\% & $43$\% & $67$\% & ${<}1$\% & $162$MHz \\
 PR & $4$ & $82$\% & $69$\% & $72$\% & $1$\% & $143$MHz \\
 \hline
 WCC & $1$ & $20$\% & $14$\% & $40$\% & $0$\% & $191$MHz \\
 WCC & $2$ & $30$\% & $23$\% & $41$\% & $0$\% & $183$MHz \\
 WCC & $4$ & $55$\% & $45$\% & $43$\% & $0$\% & $161$MHz \\
\end{tabular}

\medskip
LUTs: Look-up-tables; Regs.: Registers; BRAM: Block RAM; DSPs.: Digital signal processors; Clock freq.: Clock frequency

\end{table}

\Cref{tab:utilization} shows the different system configurations used for our experiments.
Besides the three graph problems BFS, PR, and WCC, we synthesized system variants which utilize different numbers of memory channels $p$ for $1$, $2$, and $4$ channels.
All variants have a total vertex label scratch pad size of $2^{21}$, $16$ scratch pad banks, and $8$ vertex pipelines.
All types including pointers, vertex identifiers, and vertex labels are $32$-bit unsigned integers, except PR vertex labels which are $64$-bit and consist of the degree of the vertex and its PR value.
Lastly, the depth of the reorder stage is set to $32$.
This parameterization results in a moderate resource utilization with rising look-up-table (LUT) and register (Regs.) utilization, almost constant BRAM utilization because scratch pad size is shared between the graph cores, and little clock frequency degradation.
The PR configuration has significantly higher resource utilization due to the doubled vertex label size.

{\setlength{\tabcolsep}{0.275em}
\begin{table}[bt]
\caption{Graphs used often by systems in \cref{tab:systems} (real-world graphs from \cite{LeskovecK14} and \cite{conf/aaai/RossiA15}; Graph500 generator for R-MAT)}
\label{tab:graphs}
\footnotesize
\centering
\begin{tabular}{l r r c c r r r}
 Name & $|V|$ & $|E|$ & Dir. & Degs. & $D_{avg}$ & \o & SCC \\
 \hline
 \hline
 live-journal (lj) & $4.8$M & $69.0$M & \faThumbsOUp & \includegraphics[height=1em]{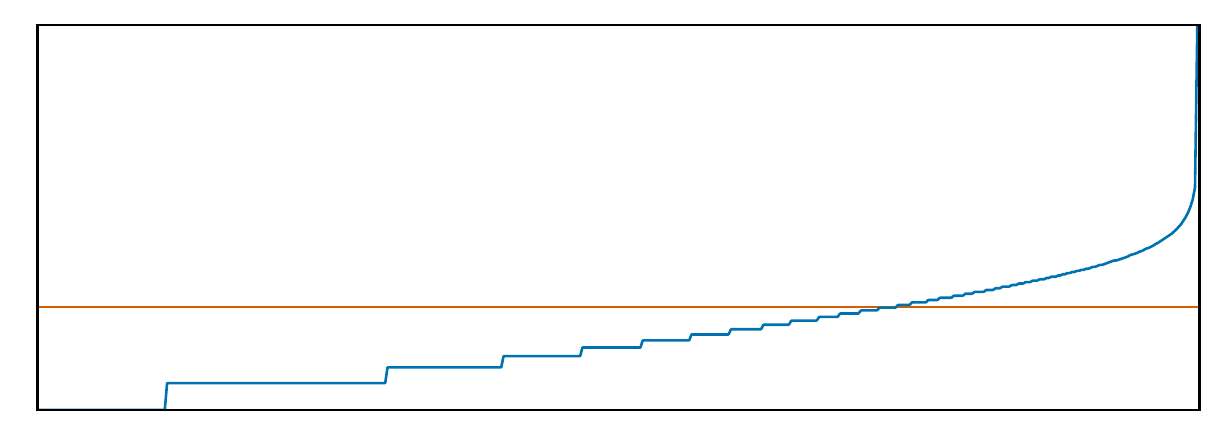} & $14.23$ & $20$ & $0.79$ \\
 orkut (or) & $3.1$M & $117.2$M & \faThumbsDown & \includegraphics[height=1em]{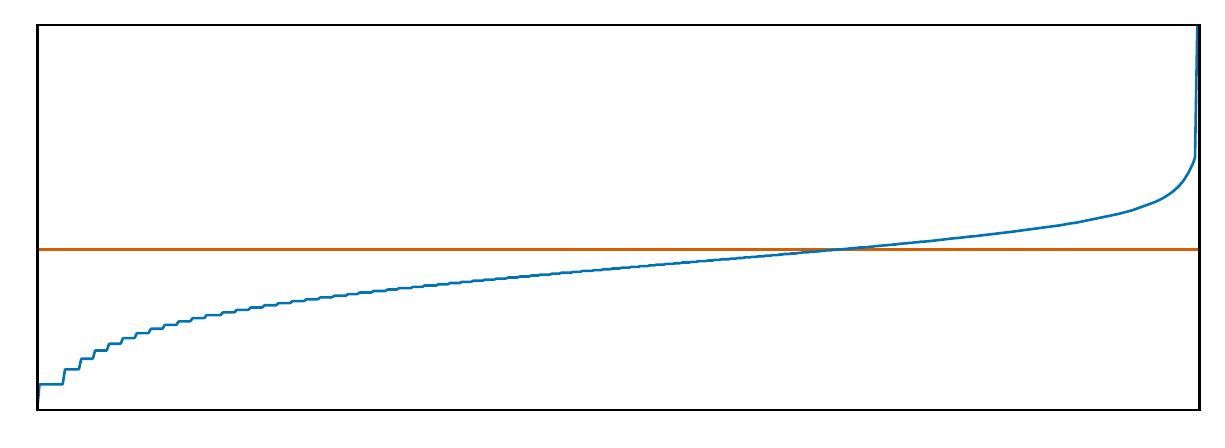} & $76.28$ & $9$ & $1.00$ \\
 wiki-talk (wt) & $2.4$M & $5.0$M & \faThumbsOUp & \includegraphics[height=1em]{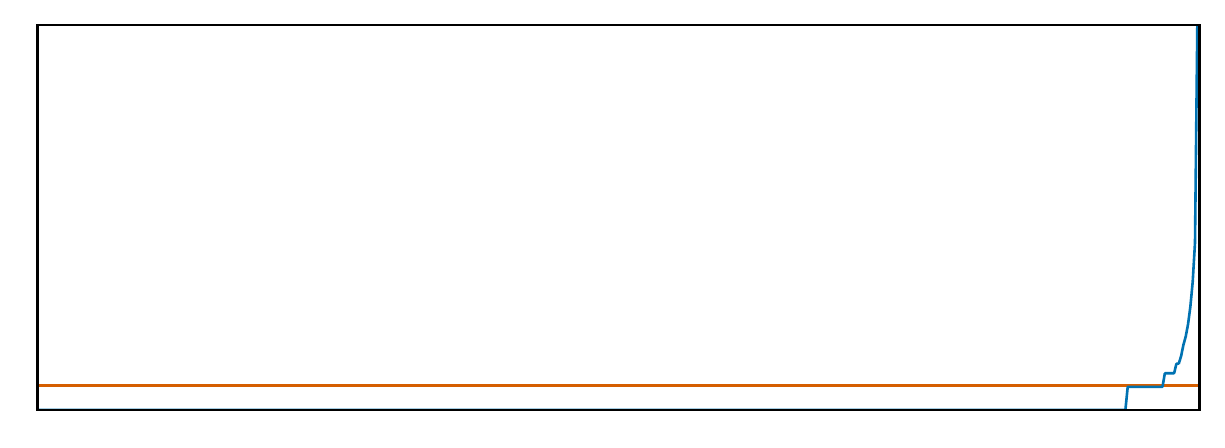} & $2.10$ & $11$ & $0.05$ \\
 pokec (pk) & $1.6$M & $30.6$M & \faThumbsDown & \includegraphics[height=1em]{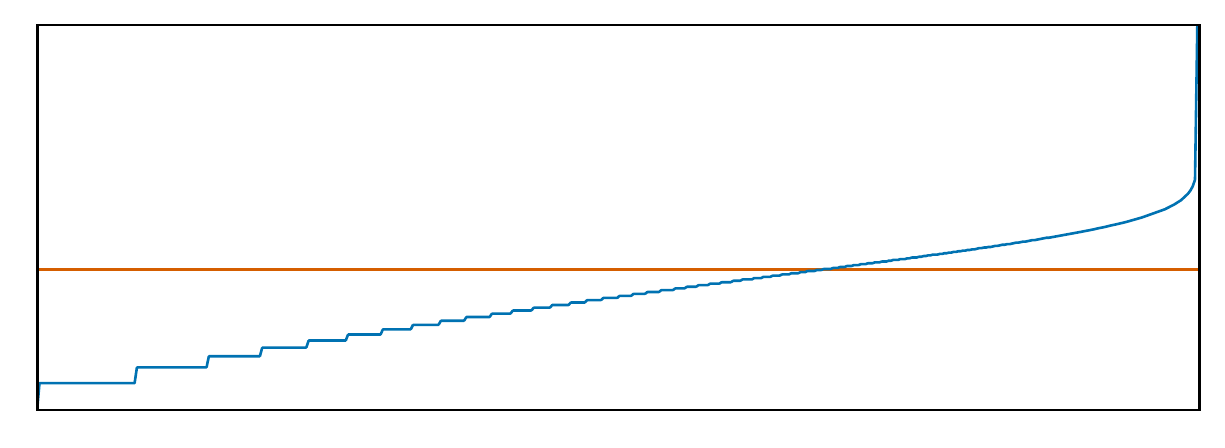} & $37.51$ & $14$ & $1.00$ \\
 youtube (yt) & $1.2$M & $3.0$M & \faThumbsDown & \includegraphics[height=1em]{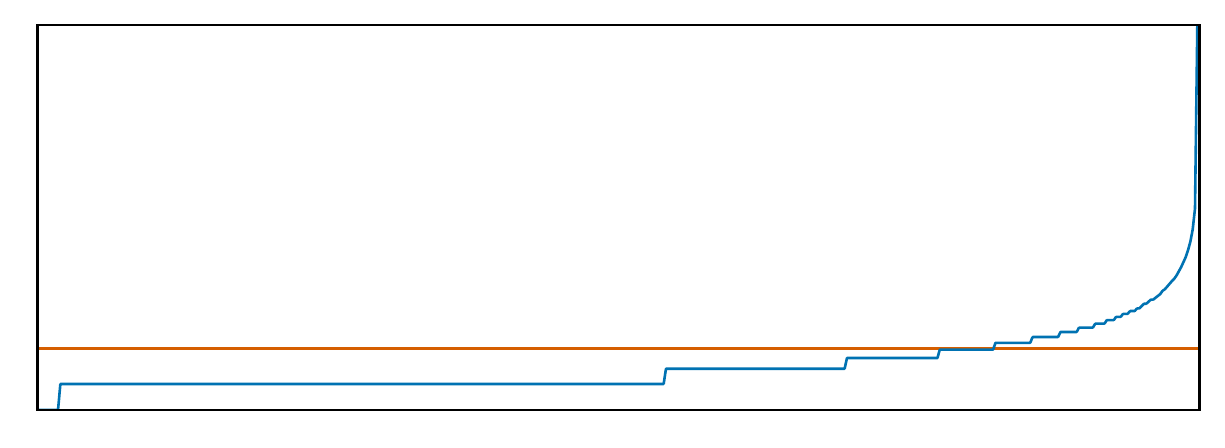} & $5.16$ & $20$ & $0.98$ \\
 dblp (db) & $426.0$K & $1.0$M & \faThumbsDown & \includegraphics[height=1em]{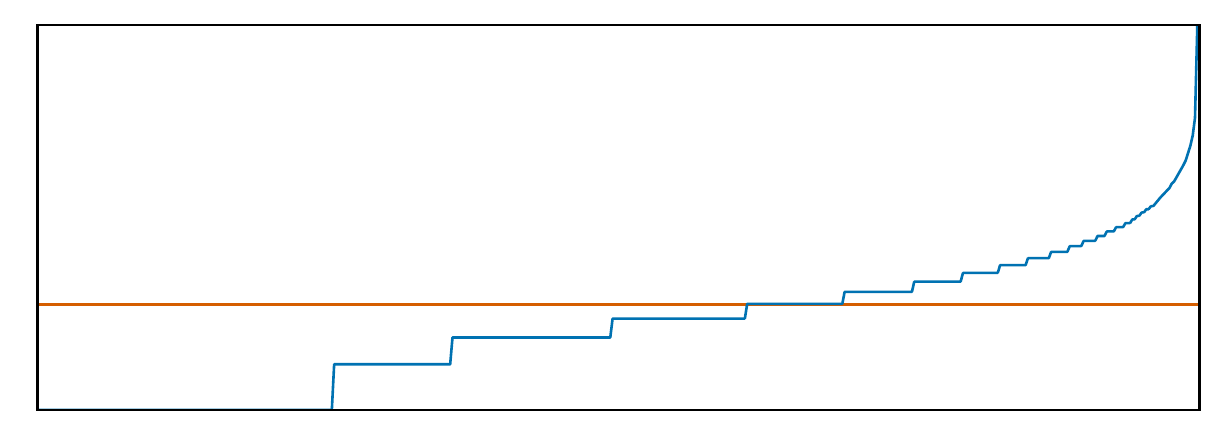} & $4.93$ & $21$ & $0.74$ \\
 slashdot (sd) & $82.2$K & $948.4$K & \faThumbsOUp & \includegraphics[height=1em]{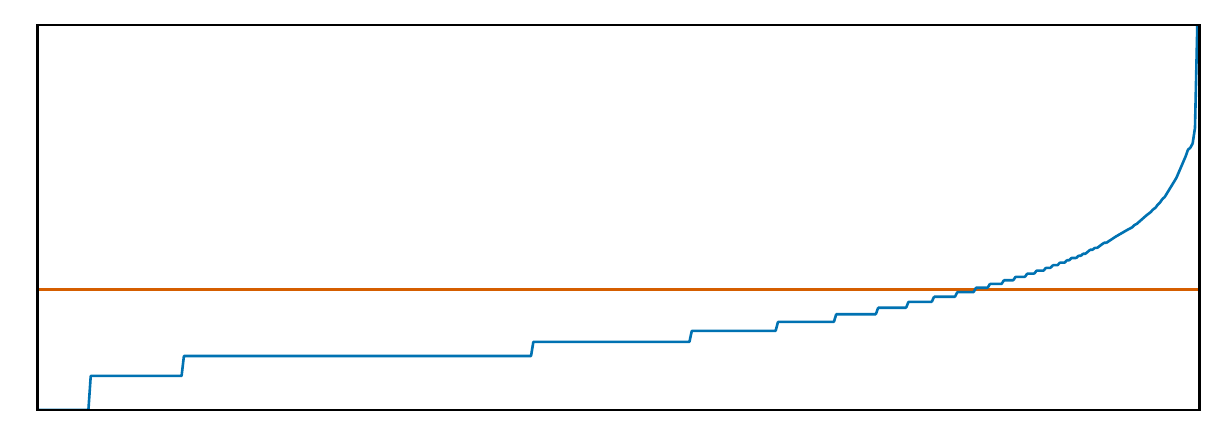} & $11.54$ & $13$ & $0.87$ \\
 mouse-gene (mg) & $45.1$K & $14.5$M & \faThumbsDown & \includegraphics[height=1em]{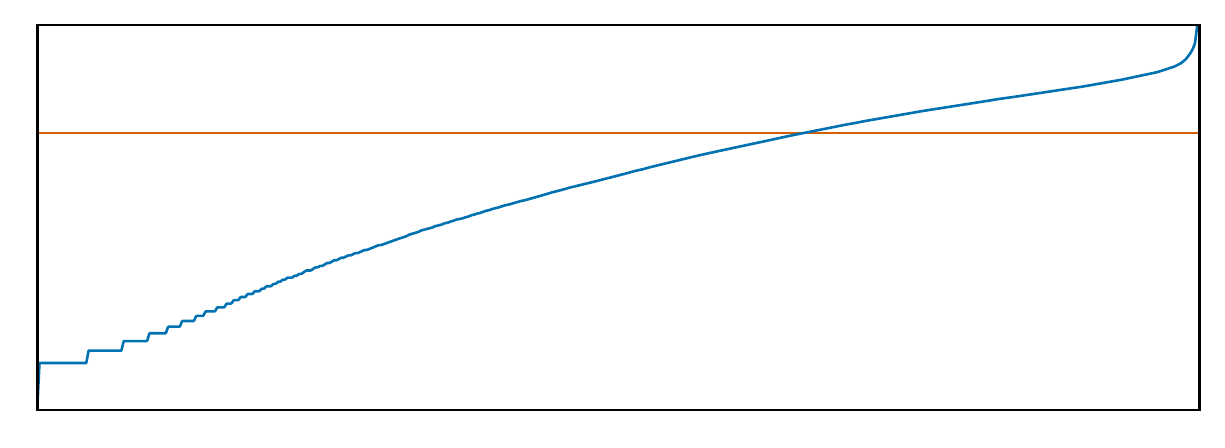} & $643.26$ & $11$ & $0.95$ \\
 \hline
 roadnet-ca (rd) & $2.0$M & $2.8$M & \faThumbsDown & \includegraphics[height=1em]{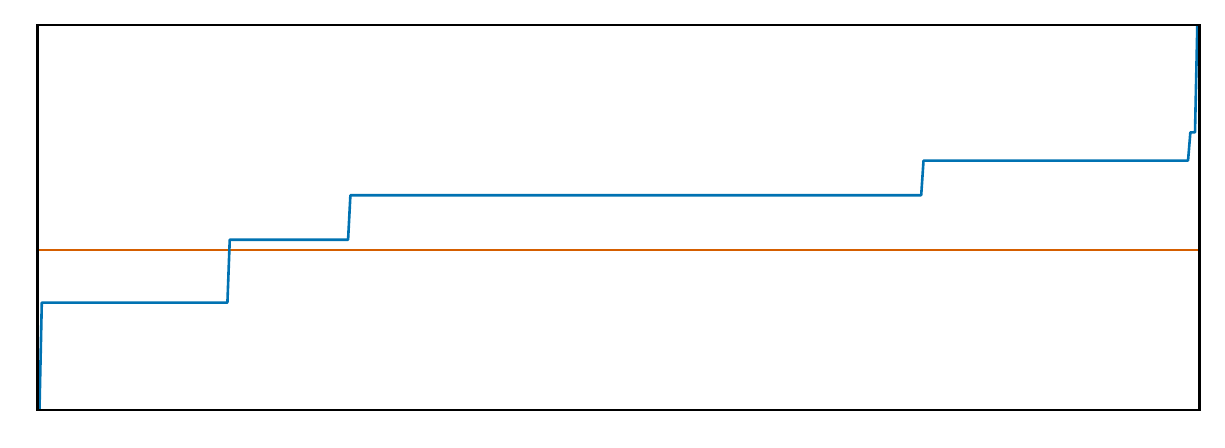} & $2.81$ & $849$ & $0.99$ \\
 top-cats (tc) & $1.8$M & $28.5$M & \faThumbsOUp & \includegraphics[height=1em]{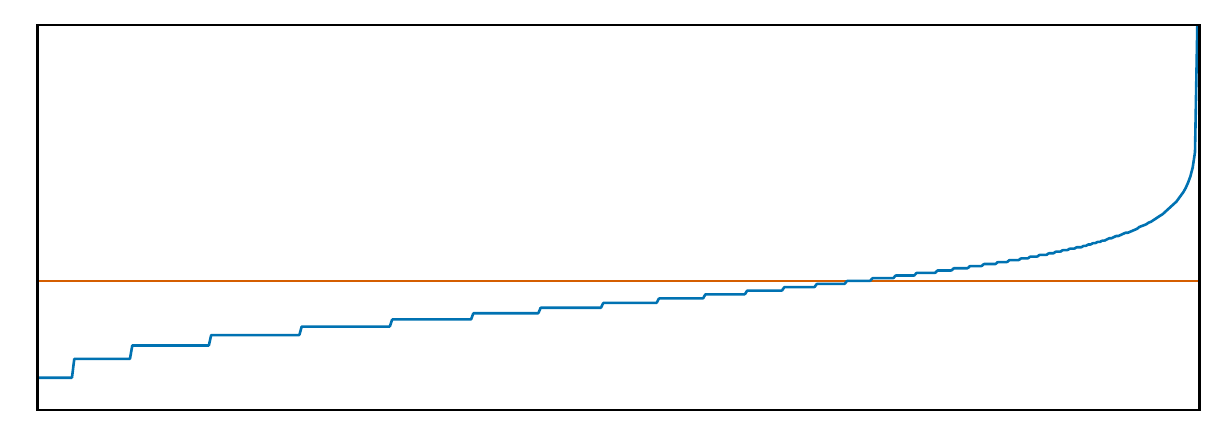} & $15.92$ & $258$ & $1.00$ \\
 berk-stan (bk) & $685.2$K & $7.6$M & \faThumbsOUp & \includegraphics[height=1em]{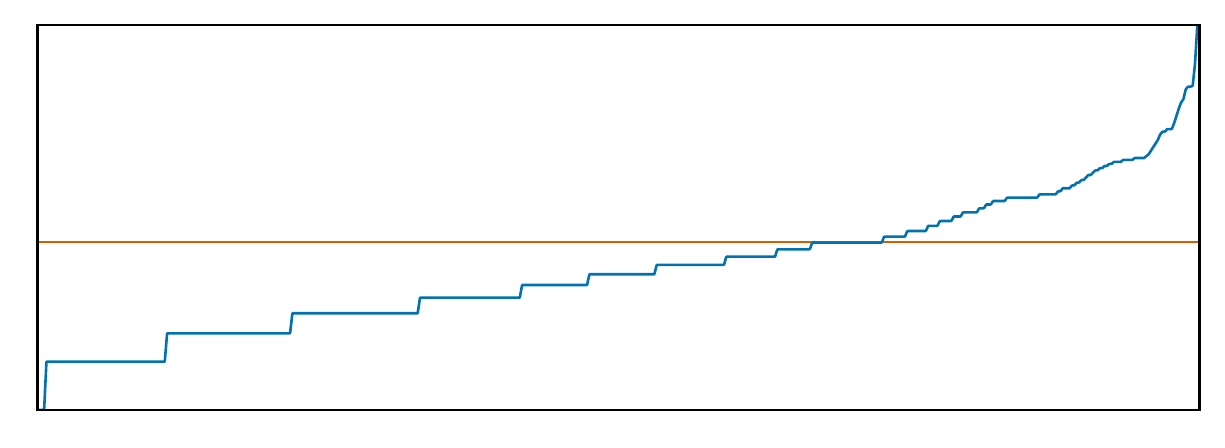} & $11.09$ & $714$ & $0.49$ \\
 \hline
 rmat-24-16 (r24) & $16.8$M & $268.4$M & \faThumbsOUp & \includegraphics[height=1em]{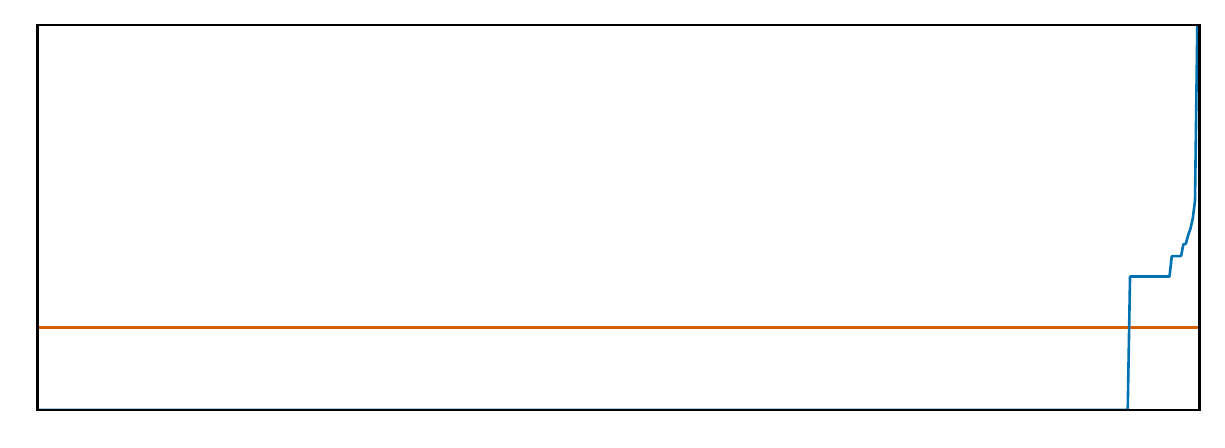} & $16.00$ & $19$ & $0.02$ \\
 rmat-21-86 (r21) & $2.1$M & $180.4$M & \faThumbsOUp & \includegraphics[height=1em]{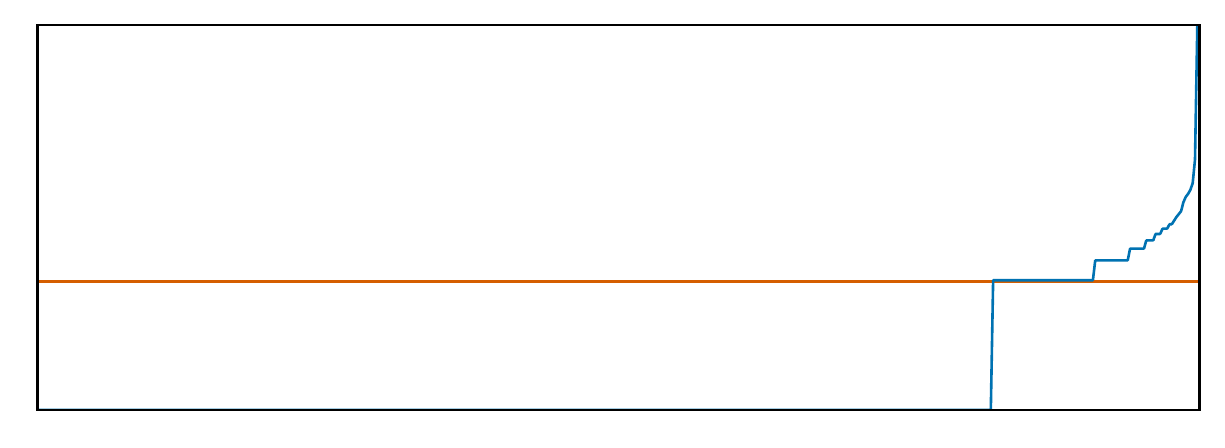} & $86.00$ & $14$ & $0.10$ \\
\end{tabular}

\medskip
Dir.: Directed; Degs.: Degree distribution on log. scale; SCC: Ratio of vertices in the largest strongly-connected component to $n$; \faThumbsOUp: yes, \faThumbsDown: no

\end{table}}
Graph data sets that are used to benchmark our system are listed in \Cref{tab:graphs}.
This selection represents the most important graphs, currently considered, found by a recent survey \cite{journals/corr/abs-2007-07595}.
Two important aspects when working with these graphs are their directedness 
and the choice of root vertices\footnote{Root vertices used: lj - $772860$; or - $1386825$; wt - $17540$; pk - $315318$; yt - $140289$; db - $9799$; sd - $30279$; mg - $20631$; rd - $1166467$; tc - $1405263$; bk - $546279$; r24 - $535262$; r21 - $74764$} (\eg for BFS or SSSP), because they can have a significant impact on performance.
We also show graph properties like degree distribution and average degree that are useful in explaining performance effects observed in the following.

\subsection{Effects of GraphScale Optimizations}
\label{sec:optimizations}
\begin{figure}[bt]
    \centering
    \includegraphics[width=\linewidth]{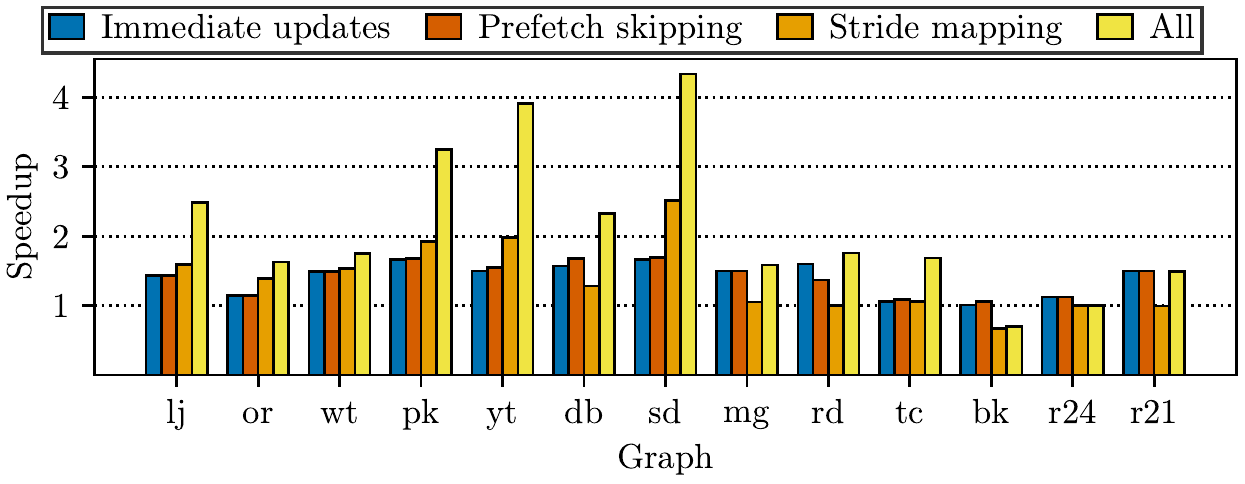}
    \vspace{-.5cm}
    \caption{Effects of GraphScale optimizations for BFS}
    \label{fig:optimizations}
\end{figure}
\cref{fig:optimizations} shows the effects of different BFS optimizations from \cref{sec:partitioning}, when applied to the base framework.
The measurements are performed on a four channel GraphScale system and normalized to measurements with all optimizations turned off.
The \emph{immediate updates} optimization ensures that updates to the vertex labels of the current partition interval are written back to the scratch pad immediately, instead of just being written back to memory.
This makes updates available earlier and leads to faster convergence for almost all graphs.
Only the berk-stan graph does not benefit from this optimization which is due to a specific combination of graph structure and selected root vertex.
The \emph{prefetch skipping} optimization skips the prefetch phase of each iteration if intermediate updates are enabled.
Hence, the prefetch skipping measurements have intermediate updates enabled.
Additionally, prefetch skipping only works on graphs with a single partition.
Prefetch skipping is a lightweight control flow optimization that sometimes leads to small performance improvements.
Lastly, \emph{stride mapping} tries to optimize partition balance.
Whenever partitions can be balanced (\eg youtube or slash-dot graphs), the performance improves most significantly.
However, in rare cases (\eg berk-stan graph) this may lead to performance degradation because with asynchronous graph processing, result convergence is dependent on vertex order and a beneficial vertex order may be shuffled by stride mapping.
From our observation, it is beneficial if high degree vertices are at the beginning of the vertex sequence for faster convergence.
In single channel measurements (not shown due to brevity), single channel performance was better without stride mapping for almost all graphs. 
This is expected because partition balance is only important between channels but not between sub-partitions.

\subsection{GraphScale Scalability}
\label{sec:scalability}
\begin{figure*}[bt]
    \centering
    \includegraphics[width=\linewidth]{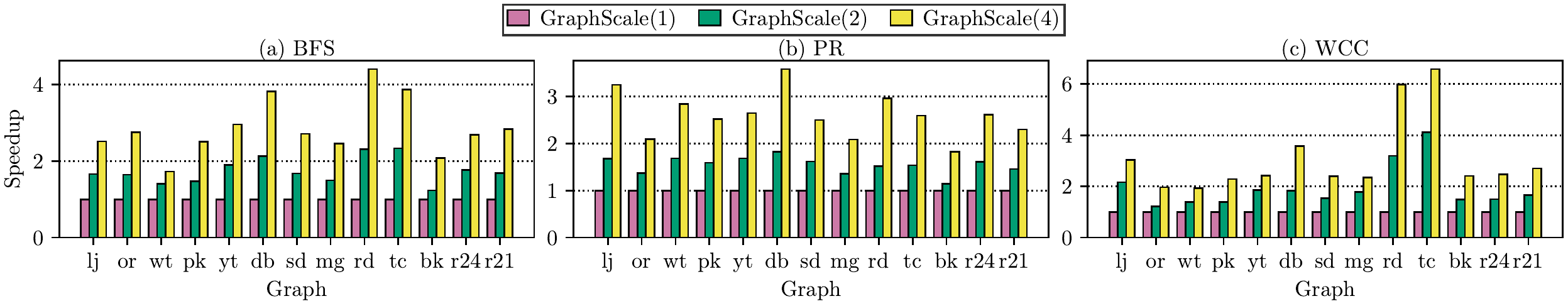}
    \vspace{-.6cm}    
    \caption{GraphScale memory channel scalability from one to four channels}
    \label{fig:scalability}
    \vspace{-.75cm}
\end{figure*}
\cref{fig:scalability} shows the scalability of GraphScale from a single-channel up to four memory channels as speed-up over the baseline of single-channel operation for BFS, PR, and WCC.
For single-channel, the stride mapping optimization is disabled.
Otherwise, all optimizations discussed in \cref{sec:optimizations} are always enabled.
The measurements show that there is some scaling overhead and speedup is dependent on the data set.
This may be due to partition balance but is mainly influenced by density (\ie average degree) of the graph for BFS. 
This can \eg be observed for the orkut, dblp, and rmat-21-86 graphs.
Two interesting exceptions are the roadnet-ca and top-cats graphs which show super-linear scaling.
This is due to stride mapping changing the vertex ordering and thus leading to convergence on the result in significantly less iterations.
Scalability speedups for WCC are similar to the BFS measurements besides the even more pronounced super-linear scaling for roadnet-ca and top-cats.

For the comparison of GraphScale against HitGraph and ThunderGP, we use the performance measure millions of traversed edge per second (MTEPS) defined by the Graph500 benchmark as $|E|/t_{exec}$ with runtime $t_{exec}$.
More is better for this performance metric.
This is different than the MTEPS* definition $|E|*i/t_{exec}$ with number of iterations $i$ used by HitGraph and ThunderGP.
MTEPS* eliminates number of iterations in favor of showing raw edge processing speed.
However, faster convergence to results due to lower number of iterations has more impact on actual runtime than usually smaller differences in raw edge processing speed.

\begin{figure}[bt]
    \centering
    \includegraphics[width=\linewidth]{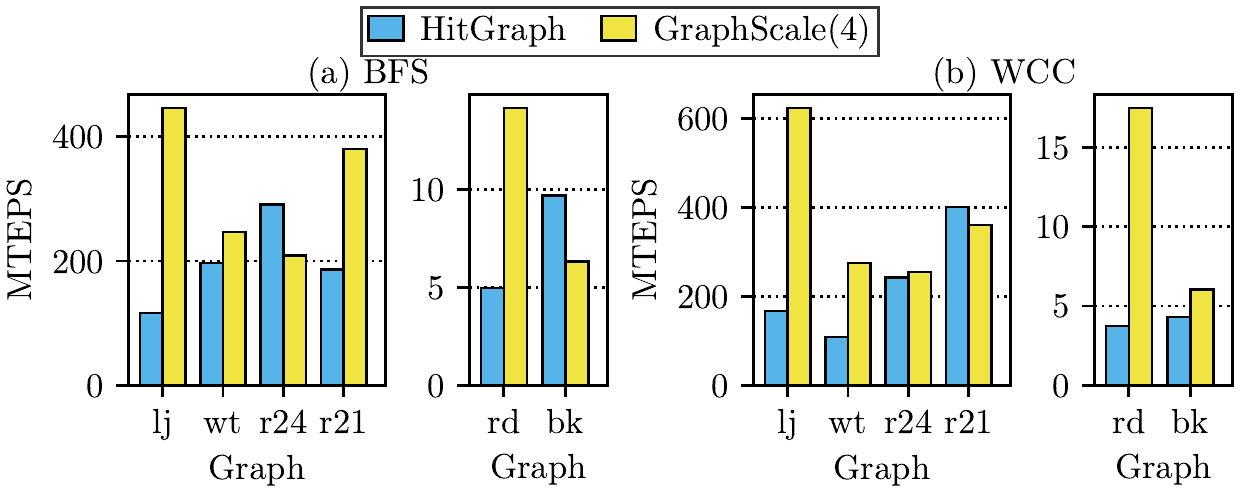}
    \vspace{-.6cm}
    \caption{Comparison of GraphScale and HitGraph}
    \label{fig:hit_compare}
    \vspace{-.6cm}
\end{figure}
\cref{fig:hit_compare} compares the four channel GraphScale system to HitGraph.
Because HitGraph does not provide BFS performance numbers, the GraphScale BFS results in \cref{fig:hit_compare}(a) are compared to single-source shortest paths results from HitGraph which has the same output for edge weights $1$.
We were not able to obtain the root vertices that were used for the HitGraph measurements thus we measure with our own.
Overall, we show an average performance improvement over HitGraph of $1.89$ for BFS and $2.38$ for WCC.
As already shown in \cref{fig:scalability}, GraphScale benefits from denser graphs like live-journal in contrast to a sparse graph like wiki-talk.
We also again observe the superior scaling of our approach for the roadnet-ca graph.
For graphs with a large vertex set like rmat-24-16, our approach requires increasingly more partitions ($9$ for rmat-24-16), introducing a lot of overhead.

\begin{figure}[bt]
    \centering
    \includegraphics[width=\linewidth]{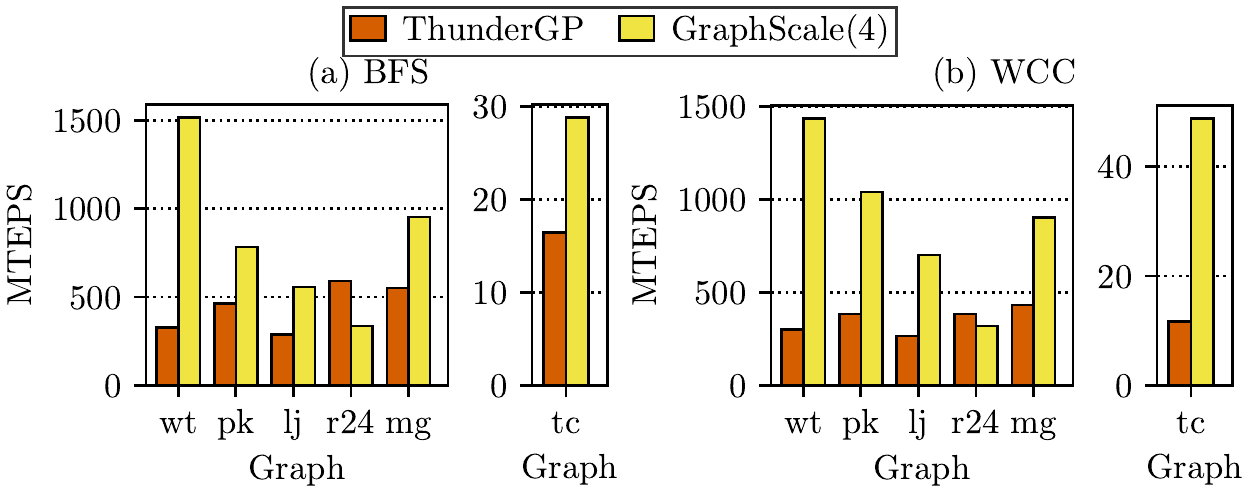}
    \vspace{-.6cm}    
    \caption{Comparison of GraphScale and ThunderGP}
    \label{fig:thu_compare}
    \vspace{-.6cm}
\end{figure}

\cref{fig:thu_compare} compares the four channel GraphScale system to ThunderGP.
For this experiment, we implemented a vertex range compression proposed by ThunderGP which removes any vertex without an outgoing edge from the graph before partitioning it.
While we apply this for the purpose of comparing the approaches on equal footing, we criticise this compression technique because it returns wrong results.
Taking BFS as an example, vertices that only have incoming edges also receive updates even though they do not propagate them.
ThunderGP uses random root vertices generated with an unseeded random generator.
Thus, we reproduce their root vertices and measure on the exact same input parameters.
Overall, we achieve a speedup over ThunderGP of $2.05$ and $2.87$ for BFS and WCC respectively.
The vertex range compression makes the wiki-talk graph much denser which our approach benefits from.
The only slowdown we observe is again for the rmat-24-16 graph due to partition overhead.

\subsection{Discussion}
\label{sec:discussion}
Overall, we observe an average speedup of $2.3$ over the two state-of-the-art graph processing accelerators HitGraph and ThunderGP with a maximum speedup of $4.77$ for BFS on the wiki-talk graph over ThunderGP, confirming the potential of scaling asynchronous graph processing on compressed data.
For optimizations, we show the importance of partition balance with stride mapping being very effective for graphs like slash-dot and youtube at the trade-off of shuffling the potentially beneficial natural vertex ordering of real world graphs.
In our scalability measurements, we observe that GraphScale performance benefits from denser graphs in general (\eg orkut and dblp graphs).
However, compared to HitGraph and ThunderGP, we observe a significant slowdown for graphs with a large vertex set, like rmat-24-16.
This results in a trade-off in the compressed data structure between less data required for dense graphs and more partitioning overhead for graphs with a large vertex set.
Partitioning overhead may be solved by further increasing scratch pad size or future work in exploring other compression techniques.
In particular, compressing the pointer array of the CSR could benefit processing on larger graphs because it gets increasingly sparser with higher partition count.

\section{Conclusion}
\label{sec:conclusion}
We propose GraphScale, an FPGA-based, scalable graph processing framework.
GraphScale is inspired by the insight that current graph processing accelerators do not combine compressed graph representations, asynchronous graph processing, and scalability to multiple memory channels in one design.
For the first time, we show the potential of such a design by defining a scalable two level crossbar and a two-dimensional graph partitioning scheme that enable all three accelerator properties in one system.
Our experimental performance evaluation shows scalability and superior performance of GraphScale compared to state-of-the-art graph processing accelerators such as HitGraph and ThunderGP with an average speedup of $2.3$ and up to $4.77$ on dense graphs.

In future work, we will further explore partitioning schemes to improve partition balance (\eg as in \cite{conf/iiswc/BalajiL18}) and decrease overhead for large graphs.
The promising results on four-channel DDR4 memory suggest an application of GraphScale to modern multi-channel memory like HBM.

\section*{Acknowledgements} We thank Intel\textsuperscript{\textregistered} Corp. for the hardware and Mehdi Moghaddamfar for valuable suggestions on the implementation.

\IEEEtriggeratref{11}

\bibliographystyle{IEEEtran}
\bibliography{IEEEabrv,paper}

\end{document}